\documentstyle{mn}

\newif\ifAMStwofonts

\def\ea{\it et al. \rm}

\ifoldfss
  \ifCUPmtlplainloaded \else
    \NewTextAlphabet{textbfit} {cmbxti10} {}
    \NewTextAlphabet{textbfss} {cmssbx10} {}
    \NewMathAlphabet{mathbfit} {cmbxti10} {} 
    \NewMathAlphabet{mathbfss} {cmssbx10} {} 
  \fi
  \ifAMStwofonts
    \ifCUPmtlplainloaded \else
      \NewSymbolFont{upmath} {eurm10}
      \NewSymbolFont{AMSa} {msam10}
      \NewMathSymbol{\upi}     {0}{upmath}{19}
      \NewMathSymbol{\umu}     {0}{upmath}{16}
      \NewMathSymbol{\upartial}{0}{upmath}{40}
      \NewMathSymbol{\leqslant}{3}{AMSa}{36}
      \NewMathSymbol{\geqslant}{3}{AMSa}{3E}

    \fi
  \fi
\fi 

\ifnfssone
  \newmathalphabet{\mathit}
  \addtoversion{normal}{\mathit}{cmr}{m}{it}
  \addtoversion{bold}{\mathit}{cmr}{bx}{it}
  \newmathalphabet{\mathbfit} 
  \addtoversion{normal}{\mathbfit}{cmr}{bx}{it}
  \addtoversion{bold}{\mathbfit}{cmr}{bx}{it}
  \newmathalphabet{\mathbfss} 
  \addtoversion{normal}{\mathbfss}{cmss}{bx}{n}
  \addtoversion{bold}{\mathbfss}{cmss}{bx}{n}
  \ifAMStwofonts
    \ifCUPmtlplainloaded \else
      \UseAMStwoboldmath
      \makeatletter
      \new@mathgroup\upmath@group
      \define@mathgroup\mv@normal\upmath@group{eur}{m}{n}
      \define@mathgroup\mv@bold\upmath@group{eur}{b}{n}
      \edef\UPM{\hexnumber\upmath@group}
      \new@mathgroup\amsa@group
      \define@mathgroup\mv@normal\amsa@group{msa}{m}{n}
      \define@mathgroup\mv@bold\amsa@group{msa}{m}{n}
      \edef\AMSa{\hexnumber\amsa@group}
      \makeatother
      \mathchardef\upi="0\UPM19
      \mathchardef\umu="0\UPM16
      \mathchardef\upartial="0\UPM40
      \mathchardef\leqslant="3\AMSa36
      \mathchardef\geqslant="3\AMSa3E
    \fi
  \fi
\fi 

\ifnfsstwo
  \DeclareMathAlphabet{\mathbfit}{OT1}{cmr}{bx}{it}
  \SetMathAlphabet\mathbfit{bold}{OT1}{cmr}{bx}{it}
  \DeclareMathAlphabet{\mathbfss}{OT1}{cmss}{bx}{n}
  \SetMathAlphabet\mathbfss{bold}{OT1}{cmss}{bx}{n}
  \ifAMStwofonts
    \ifCUPmtlplainloaded \else
      \DeclareSymbolFont{UPM}{U}{eur}{m}{n}
      \SetSymbolFont{UPM}{bold}{U}{eur}{b}{n}
      \DeclareSymbolFont{AMSa}{U}{msa}{m}{n}
      \DeclareMathSymbol{\upi}{0}{UPM}{"19}
      \DeclareMathSymbol{\umu}{0}{UPM}{"16}
      \DeclareMathSymbol{\upartial}{0}{UPM}{"40}
      \DeqsleclareMathSymbol{\lant}{3}{AMSa}{"36}
      \DeclareMathSymbol{\geqslant}{3}{AMSa}{"3E}
    \fi
  \fi
\fi 

\ifCUPmtlplainloaded \else
  \ifAMStwofonts \else 
    \def\upi{\pi}
    \def\umu{\mu}
   \def\upartial{\partial}
  \fi
\fi

\title{The Component Star Masses in RW Tri}
\author[T. Poole et al.]
       {T. Poole,$^1$\thanks{tsp@mssl.ucl.ac.uk} K. O. Mason,$^1$
G. Ramsay,$^1$ J. E. Drew$^2$, R. C. Smith$^3$\\
	$^1$Mullard Space Science Laboratory, University College London,
Holmbury St. Mary, Dorking, Surrey RH5 6NT\\
	$^2$Imperial College of Science, Technology \& Medicine, Blackett
Laboratory, Prince Consort Road, London SW7 2BW\\
	$^3$Astronomy Centre, School of Chemistry Physics and Environmental
Science, University of Sussex, Brighton, BN1 9QJ\\}
\date{in accepted form 2002 November 27}

\pagerange{\pageref{firstpage}--\pageref{lastpage}}
\pubyear{2002}

\begin{document}

\maketitle

\label{firstpage}

\begin{abstract}
We use time-resolved spectra of the cataclysmic variable RW Tri in the I and K-band to 
determine the orbital velocity of the
secondary star using skew mapping and cross correlation techniques
respectively. We find radial velocity amplitudes of $250\pm47$ km/s in the I-band, 
and $221\pm29$ km/s in the K-band. We also determine the rotational velocity of
the secondary star using the K-band data and find a $V_{rot}\sin i$ of $120\pm20$km/s.
A combination of these results coupled with an estimate of the effect of heating
on the secondary star suggests a mass ratio $M_2/M_1$ in the range $0.6-1.1$;
the mass ratio range with no correction for heating is $0.5 - 0.8$.
These lead to most likely estimates of the primary and secondary star masses 
in the range $0.4-0.7M_{\odot}$ 
and $0.3-0.4M_{\odot}$ respectively. Further refinement of the stellar masses
is 
hampered by uncertain knowledge of the white dwarf orbital velocity,  
and we discuss evidence that at least some estimates of the white dwarf velocity
are 
contaminated by non-orbital components. 
\end{abstract}

\begin{keywords}
techniques: radial velocities -- binaries: eclipsing -- stars: individual: RW
Tri -- novae, cataclysmic variables -- infrared: stars.
\end{keywords}

\section{Introduction}
RW Tri is an eclipsing nova-like cataclysmic variable (CV) system with an orbital period of
5 hours 34 mins.  Nova-like systems contain a non-magnetic white dwarf primary and a
late type (K-M) secondary star that fills its Roche lobe, transferring mass to
the white dwarf via an accretion disc. 

Since first being observed as an eclipsing variable (Protitch 1937)
multi-wavelength studies of RW Tri have provided much insight into the system,
but have yet to yield an accurate measurement of the masses of the two
component stars. By observing the velocities of the component stars, their masses can
be calculated using Kepler's laws. A number of authors have attempted 
to measure the
radial velocity of the white
dwarf (e.g. Kaitchuck \ea 1983, Still \ea 1995 and
Mason \ea 2002), but thus far little work has been done on the secondary star.  

Measurement of the secondary star velocity is hampered by the large contrast in optical brightness between it and the accretion disk. 
Weak  secondary star absorption features were detected in the I-band spectrum of RW Tri by Friend \ea (1988), and later used by Smith \ea
(1993) to estimate the secondary star velocity as $\sim250$km/s using a `skew-mapping' technique (cf. Smith \ea 1998a). Dhillon \ea (2000) have detected secondary star features in low-resolution K-band spectra of RW~Tri. The relative contribution of the secondary star is higher in the K-band and the equivalent width of the absorption features greater as a consequence. 

In this paper we report phase-resolved spectral observations of RW Tri in the
K-band, covering the binary orbit. We use these observations to 
determine the radial velocity of the secondary star in RW Tri.
We also re-analyse the original I-band spectra of Smith \ea (1993).  Based on these results we revisit the mass ratio
and the masses of the component stars in RW~Tri and discuss their implications.

\section{I-band Data}
We have re-analysed the RW~Tri near I-band data of Smith \ea (1993) using the back-projection routines in the software package 
MOLLY written by Tom Marsh (see also Vande Putte \ea 2002), combined with a
Monte Carlo error analysis. We use the method of `skew-mapping' (Smith \ea
1993) in which the orbital phase-resolved data are corrected for a trial
orbital velocity amplitude and phase, and compared to the spectrum of a
standard `template' star. The degree of correlation between the summed
corrected spectrum and the template is computed for a grid of velocity
amplitudes and phases to produce the skew map.  This method is useful when the
spectral features of the secondary are too weak to produce significant 
cross correlation peaks using
individual spectra.

 The I-band spectra were obtained using the Isaac
Newton Telescope (INT) on La Palma, and the 831R grating, giving a wavelength
coverage of 7700\AA\ - 8300\AA.  This range includes the Na I
absorption doublet feature ($\lambda$8183.3\AA, $\lambda$8194.8\AA) that is expected in the spectrum of a late-type star. A total of 27 spectra were taken over two
nights spanning the full orbital cycle: 19
spectra were taken on the 2$^{nd}$ September 1985, and 8 spectra were taken on
3$^{rd}$ September 1985.  The ephemeris of Robinson \ea (1991) was used to phase the spectra. A range of 7 template stars of spectral type K5-M1.5
were taken from Martin (1988), the details of which are given in Table~\ref{t1}. Template and RW Tri
spectra were normalised by dividing by a low-order spline fit to the respective continuum. 

\begin{table}
\caption{Template stars from Martin (1988).}
\begin{tabular}{ccc}
\hline
Name & Color Index & Spectral type \\
\hline
Gl 653 & 0.49 & K5\\
Gl 717 & 0.54 & K7\\
Gl 673 & 0.60 & K7\\
Gl 488 & 0.67 & M0-\\
Gl 383 & 0.70 & M0\\
Gl 281 & 0.71 & M0\\
Gl 908 & 0.87 & M1.5\\
\hline
\end{tabular}
\label{t1}
\end{table}

A range of skew maps with total systemic velocities (RW Tri $+$ template) from
$-50$km/s to $50$km/s, were produced for each template.   
A mask between 7760\AA\ and 7800\AA\ was used to
remove the accretion disc absorption feature of neutral oxygen at 7774\AA\
(Friend \ea 1988).  The skew map with the strongest peak was found using
the M0 template Gl 281, with a RW Tri systemic velocity of -13 km/s and secondary
velocity amplitude of 250km/s.    Patterson (1984) estimated the secondary star of RW
Tri to be of spectral type M0, based on the empirical zero-age main-sequence (ZAMS)
mass-radius relation. Our result is also consistent with the statistical prediction of K9
$\pm$ 3 by Smith \&
Dhillon (1998b). The other M0 template (Gl 383 - Table~\ref{t1}) and the M0- template
(Gl 488 - Table~\ref{t1}), produce skew maps
with $K_2$ of 259 km/s and 255 km/s respectively, consistent with the result for
template Gl 281.

\begin{figure}
\centering
\vspace*{6.5cm}
\leavevmode
\includegraphics{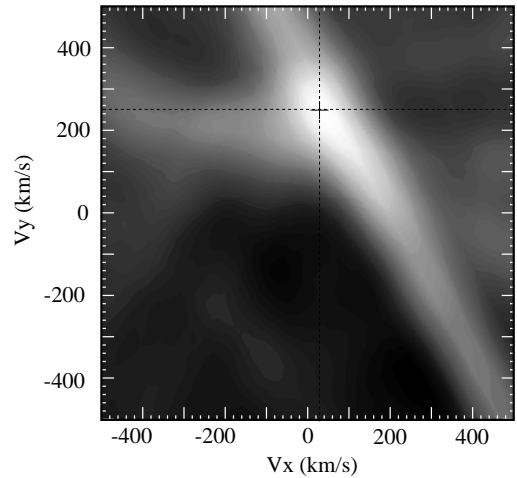}
\caption{Best fit skew map of RW Tri with template Gl 281, with a systemic
velocity of -13 km/s.  The white of the grey scale represents the strongest cross correlation. The peak of $250\pm47$km/s on the map is represented by +.}
\label{f1}
\end{figure}

A Monte Carlo simulation was
used to calculate the error on the skew map results. The noise level on the normalised RW Tri spectra was used as a seed to generate
normally-distributed random numbers that were added to the RW Tri spectra.  In
this way 150 skew maps were produced and the distribution of peak velocity values was measured to derive an error.  Each trial was weighted with the line integral intensity of its
respective map in computing the standard deviation of the resulting
distribution.  In this way, we determine that the best fit skew map
(Figure~\ref{f1}) has a secondary velocity amplitude of $K_2=250\pm47$
km/s.  

Vande Putte \ea (2002) have obtained results using the same set of RW Tri data
but a different set of templates ranging from K5-M4.5.  They find a
best fit using the M1 template Gl 514 and $K_2=263\pm30$ km/s with a systemic
velocity of $-11 \pm 15$km/s agreeing with our results within error.
The differences between their value of $K_2$ and ours (which is within the
respective errors) can be ascribed to the different set of templates used, and
is a measure of the systematic uncertainty. 

\section{K-Band Data}
The secondary star only contributes $\sim$10\% of the flux in the I-band, compared
to an estimated $\sim$65\% in the K-band (Dhillon \ea 2000), so
we would expect to see secondary star absorption features more easily at longer wavelengths.  Absorption
features due to the secondary star have been observed in the K-band spectrum
of RW Tri by Dhillon \ea (2000) at low spectral resolution, which motivated us to investigate whether we could use these features to measure the secondary star velocity directly using a standard cross correlation technique.

\subsection{Observations}
Observations of RW Tri were made in the K-band on 
the 6$^{th}$ and 7$^{th}$ August 2000 using CGS4 on the
United Kingdom Infra-Red Telescope (UKIRT). The 150 line/mm
grating gives us a spectral resolution of 100km/s at
2.2$\mu$m, and a wavelength range of $2.200\mu$m to $2.275\mu$m, enabling us
to observe both the Na I and Ca I absorption lines recorded by Dhillon \ea (2000) in RW Tri. The
telescope was nodded up and down the slit taking spectra at different detector
positions to facilitate the removal of sky background. Summing the data from the telescope nodding positions gave  a total
exposure time of 600s for each RW Tri spectrum.  This exposure time is
equivalent to 0.03 orbital cycles which is short
enough to prevent any significant smearing due to orbital effects.

A total of 35 spectra were taken of RW Tri over the two nights: 15 spectra on
the first night, and 20 spectra on the second.  Ten template stars in the range
of spectral type K7-M2 were observed over the two nights
(Table~\ref{t2}). Bright A-type stars of known broadband magnitudes and arc spectra were
also observed through both nights, enabling us to flux and wavelength calibrate
respectively.  The bright A-stars also allowed us to remove telluric lines.

\begin{table}
\caption{Template stars observed in the K-band.}
\begin{tabular}{cccc}
\hline
Name & Spectral  & Observation & Exposure\\
 &type &date & Time (s)\\	
\hline
Gl 397 & K5 & 7/9/00 & 5 \\
Gl 3478 & K7 & 7/9/00 & 20 \\
Gl 334 & K7 & 6/9/00 & 10 \\
Gl 182 & M0 & 6/9/00 & 20 \\
Gl 281 & M0 & 7/9/00 & 12 \\
Gl 383 & M0 & 7/9/00 & 14 \\
Gl 212 & M0.5 & 6/9/00 & 10 \\
Gl 390 & M1 & 7/9/00 &  10 \\
Gl 382 & M1.5 & 7/9/00 & 7 \\
Gl 393 & M2 & 7/9/00 & 7 \\
\hline
\end{tabular}
\label{t2}
\end{table}

\subsection{Data Reduction}
Real time data reduction was undertaken at the telescope using the ORAC-DR
package developed at the Joint Astronomy Centre. Bad pixels were masked and a flat field was applied to
remove pixel-to-pixel variations across the array, and a bias frame was
subtracted. As a result of nodding the telescope along the slit, the night sky
spectrum could be accurately removed and the spectra coadded.

Bad pixels were found close to the Na I doublet feature at 22063\AA\ and 22101\AA\ on night 1, and 22078\AA\ and 22116\AA\ on night
2. The difference between the apparent wavelength of the bad pixels between the nights
is due to a slight shift in the grating angle. These bad pixels were removed by interpolation.

Further data reduction was carried out using the FIGARO package. Spectra were extracted, and wavelength
calibrated using the rest wavelength of arc spectra, and telluric lines were
removed using the observations of A-stars.  The resulting RW Tri and template spectra were
flux calibrated using the A-stars, and then smoothed and rebinned on to
a linear wavelength scale. The ephemeris of Robinson \ea (1991) was used to phase the data. 

\begin{figure}
\centering
\vspace*{9.0cm}
\leavevmode
\includegraphics{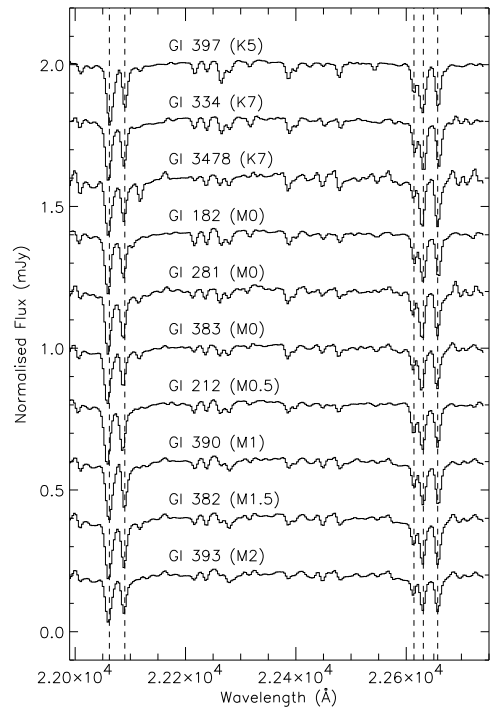}
\caption{The normalised template spectra observed using UKIRT.  The double
dashed lines
on the left of the plot represent the rest wavelengths of the Na I doublet $\lambda$22062\AA\
$\lambda$22090\AA\, and the triple dashed lines on the right represent the rest
wavelengths
of the Ca I triplet
$\lambda$22614\AA\, $\lambda$22631\AA\, $\lambda$22657\AA. The spectra are
offset vertically by 0.2 normalised flux (mJy). }
\label{f2}
\end{figure}

\begin{figure*}
\centering
\vspace*{21.5cm}
\leavevmode
\includegraphics{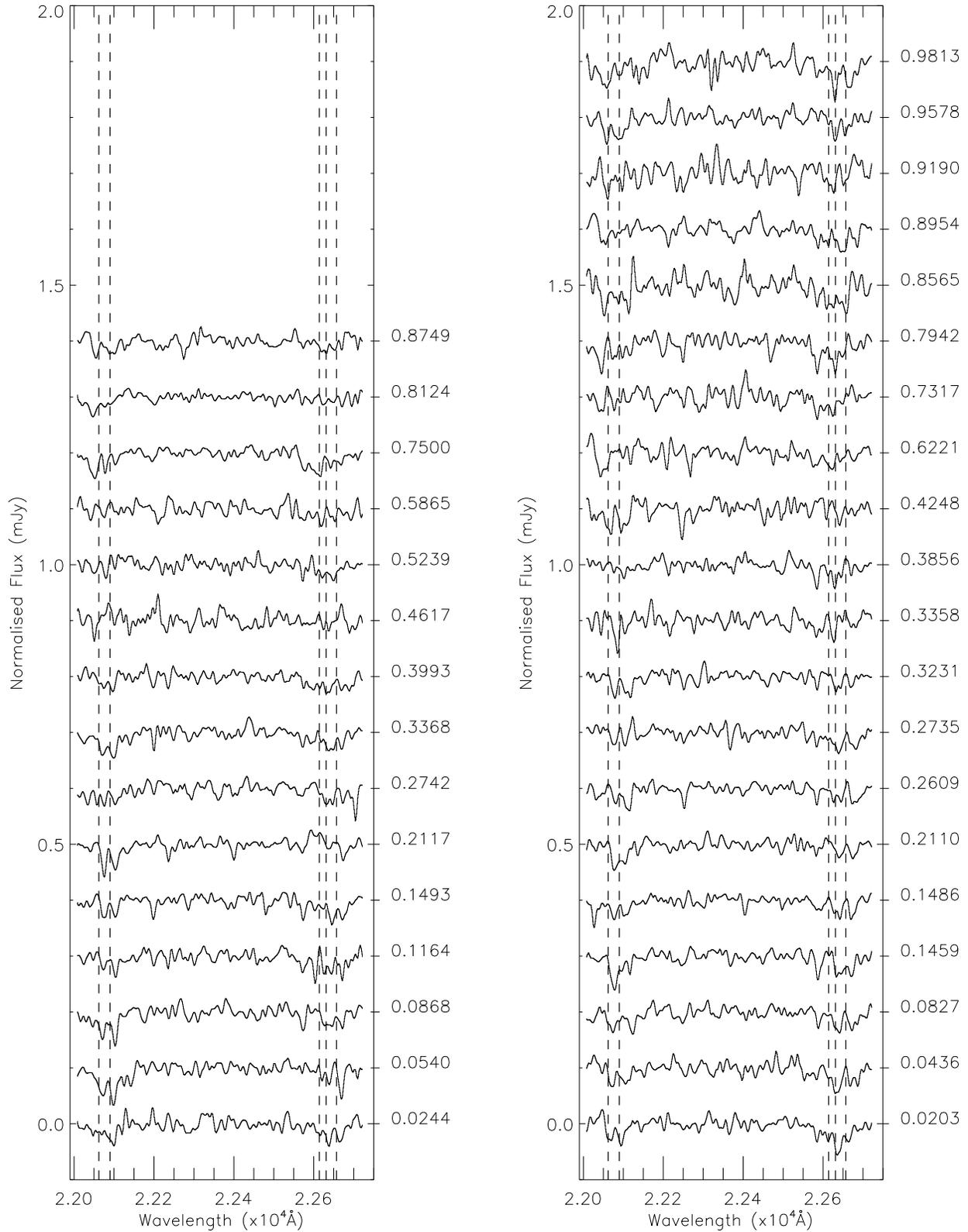}
\caption{The left plot shows the smoothed normalised RW Tri spectra of the
first night, and the right plot shows the smoothed normalised RW Tri spectra of
the second night. The orbital phase is printed on the right hand side axes of
both plots.  The dashed lines in both plots
represent the rest wavelengths of the Na I doublet (left) and Ca I triplet (right) as in Figure~\ref{f2}. Each
spectrum is vertically offset by 0.1 normalised flux from its neighbour.}
\label{f3}
\end{figure*}

\begin{figure}
\centering
\vspace*{6.0cm}
\leavevmode
\includegraphics{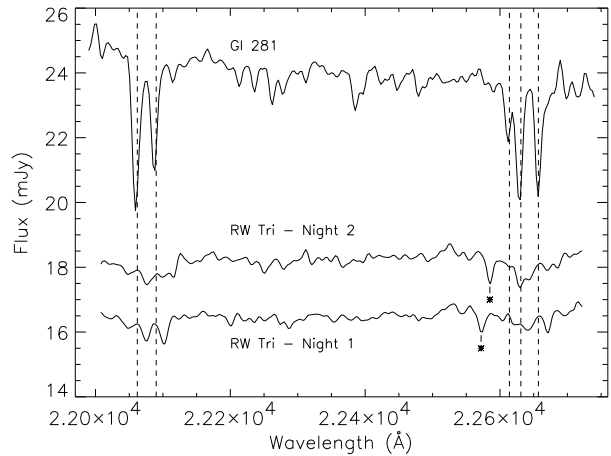}
\caption{Average RW Tri spectra for each night. The M0
template Gl 281 is flux scaled by a factor 1/170 and offset by 1.5mJy, and the
average night 2 data is offset by 1.2mJy. The absorption feature marked by $\ast$ is due to bad pixels.  As in
Figure~\ref{f2} the dashed lines
represent the rest wavelengths of the Na I doublet (left), and the Ca I triplet
(right).}
\label{f4}
\end{figure}

Figure~\ref{f2} shows the normalised template spectra in order of spectral
type.  The
Na I doublet and Ca I triplet absorption features can be clearly seen showing
the quality of the data.  There are no large differences in the
depths of the Ca I and Na I lines between the different spectral types in this
wavelength band, making it hard to distinguish between them. 
Figure~\ref{f3} shows the individual spectra of RW~Tri distributed in orbital phase.
The Na I and Ca I absorption features appear to shift from spectrum to spectrum on each night.
These
features will be discussed further in Section 3.4.  RW Tri
was brighter on the second night.  Figure~\ref{f4} shows the average spectrum of RW
Tri for each night with the scaled M0 template Gl 281.  

\subsection{The Secondary Star Velocity}
\begin{figure}
\centering
\vspace*{6cm}
\leavevmode
\includegraphics{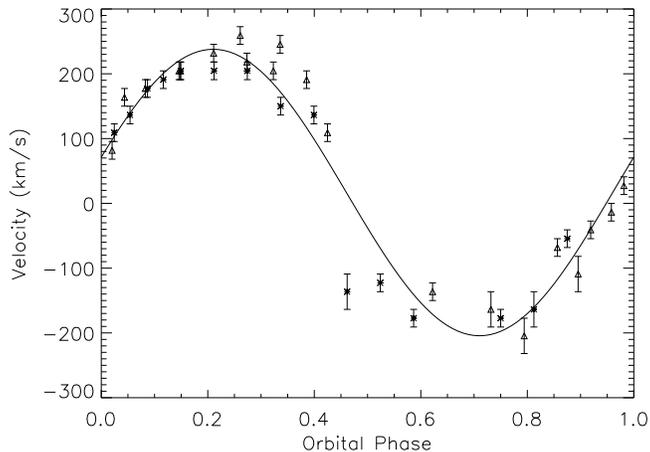}
\caption{The best fit radial velocity curve of the two nights.
The symbols $\ast$ and $\bigtriangleup$ represents the data points from
night 1 and 2 respectively.  Each point is the mean of the lags derived from the
various stellar templates used (Table~\ref{t2}) and the error bars are the standard deviation
of the results from the different templates, and therefore express the systematic
error on the data due to template choice. The
sinusoidal line represents the best fit velocity curve derived by least-squares fitting
to the data.}
\label{f5}
\end{figure}

The velocity of the secondary star in RW~Tri was investigated by cross correlating the 
spectra of RW~Tri with template spectra.  Both sets of spectra
were normalised in the same way as described for the
I-band data (Section 2), and rebinned on to the same
wavelength scale.  Each RW Tri spectrum was cross correlated with each
template spectrum in turn.  The position of the cross
correlation peak was then plotted against orbital phase to produce the
velocity curve of the secondary star. As in skew-mapping (Section 2), the
spectral template that yields the strongest cross correlation peaks is 
the best fit to the data.
All the cross correlation templates gave similar peak values, 
which is what we expect as there is little variation in
the relative strengths of the  Na I and Ca I lines 
over the range of late-type stars that 
we used (cf Figure~\ref{f2}).  The
best fit radial velocity curve of the average cross correlation lags using
all the template data is illustrated in Figure~\ref{f5}. 

This velocity curve can be parameterised as,
\begin{equation}
V=\gamma+K\sin(2\pi(\Phi-\Phi_o)),
\end{equation}
where {\it V} is the radial velocity, $\gamma$ is the systemic velocity, $K$
is the orbital velocity amplitude, $\Phi$ is the binary orbital
phase defined by the ephemeris of Robinson \ea (1991), and $\Phi_o$ is the
orbital phase of blue to red zero crossing (inferior conjunction of the
secondary star). A sinusoidal curve was fitted to the velocity data
using the Levenberg-Marquardt algorithm in IDL; by
minimising the reduced chi-squared and considering the 68.3 \%
(1 sigma) confidence level, a best-fit velocity and error was calculated.  This gives a $K_2=221\pm29$
km/s, $\gamma=17\pm20$ km/s and $\Phi_o=-0.040\pm 0.020$ orbital
phase.  
The results derived from individual template stars are listed in Table~3.
These are all consistent within the errors, and consistent with the mean
values listed above.

\begin{table}
\caption{Secondary star orbital parameters derived from individual 
template stars.}
\begin{tabular}{cccc}
\hline
Template & $K_2$  & $\gamma$ & $\Phi_o$\\
 & (km/s) & (km/s) & (orbital Phase)\\	
\hline
Gl 397 & $229\pm47$ & $-6\pm28$ & $-0.044\pm0.031$ \\
Gl 3478 & $218\pm92$ & $18\pm65$ & $-0.037\pm0.068$ \\
Gl 334 & $218\pm38$ & $10\pm27$ & $-0.042\pm0.027$ \\
Gl 182 & $217\pm51$ & $21\pm36$ & $-0.041\pm0.037$ \\
Gl 281 & $223\pm10$ & $28\pm6$ & $-0.039\pm0.007$ \\
Gl 383 & $220\pm4$ & $36\pm3$ & $-0.042\pm0.003$ \\
Gl 212 & $219\pm90$ & $39\pm64$ & $-0.041\pm0.064$ \\
Gl 390 & $221\pm59$ & $0\pm37$ & $-0.042\pm0.037$ \\
Gl 382 & $222\pm5$ & $14\pm8$ & $-0.042\pm0.008$ \\
Gl 393 & $217\pm38$ & $17\pm27$ & $-0.041\pm0.026$ \\
\hline
\end{tabular}
\label{new}
\end{table}

\begin{figure}
\centering
\vspace*{6.0cm}
\leavevmode
\includegraphics{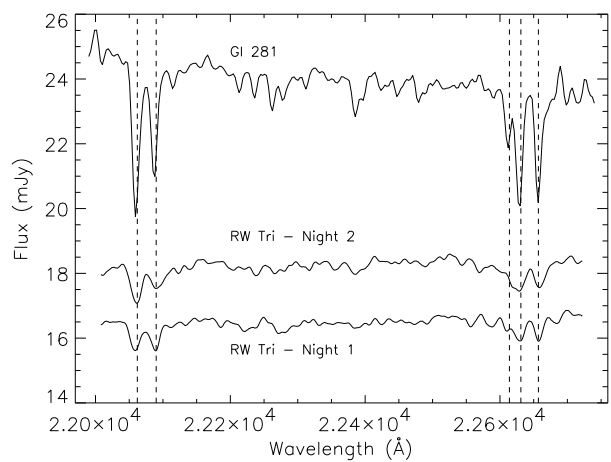}
\caption{Velocity corrected average RW Tri spectra for each night. The M0
template Gl 281 is flux scaled by a factor 1/110 and offset by 5mJy, and the
average velocity corrected night 2 data is offset by 1.2mJy. As in
Figure~\ref{f2} the dashed lines
represent the rest wavelengths of the Na I doublet (left), and the Ca I triplet
(right).}
\label{f7}
\end{figure}

In Figure~\ref{f7} we plot the velocity corrected average spectrum of RW Tri
for each night, along with the scaled M0 template Gl 281. We can clearly see the
Na I doublet and Ca I triplet in both nights.  These features are broadened
compared with the template spectra, due to the rotational velocity of the
secondary star. 

\subsection{Orbital Modulation of Secondary Star Absorption Features}
\begin{figure}
\centering
\vspace*{6.0cm}
\leavevmode
\includegraphics{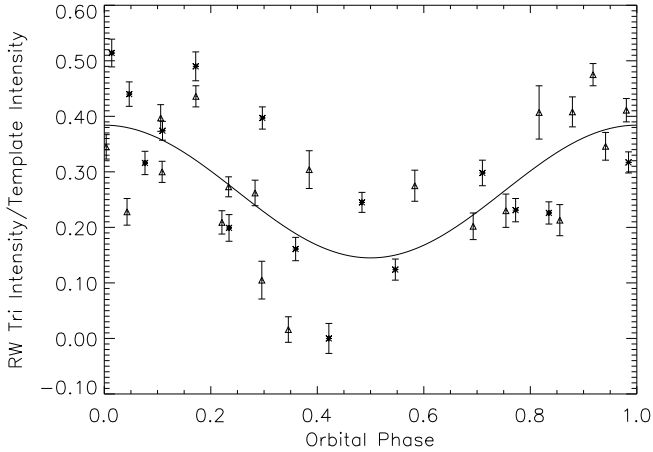}
\caption{The total flux deficit in the Na~I and Ca~I RW Tri secondary star 
absorption features, expressed as an equivalent width, divided by the corresponding 
equivalent width in the template star Gl~281 (which has the highest quality spectrum among the
templates measured, and is the best match to RW~Tri in the I-band). 
We corrected for the orbital velocity shift of RW Tri in forming these numbers. 
Night 1 is represented by  $\ast$, and night 2
is represented by $\bigtriangleup$. The solid line represents the best fit
sinusoidal curve to the data.}
\label{f8}
\end{figure}

To search for variations in the strength of the absorption features with
orbital phase, we compare the depth of the strongest features in the velocity 
corrected spectrum of RW Tri with the values observed in the
the template star spectrum.  We considered the combined regions around
the Na I absorption feature (22010\AA\ - 22140\AA) and the Ca I absorption
feature (22570\AA\ - 22700\AA), and mask out the rest of the spectrum.  
Figure~\ref{f8} shows a plot of
the ratio of RW Tri versus template star absorption feature deficit through the orbital
cycle. There is some evidence that the secondary features are strongest near
phase zero and weaker near phase 0.5.  Taken at face value, this suggests that the centroid
of the secondary features is shifted to the hemisphere of the secondary
that faces away from the disc.  Although the effect is marginal, it is in
accordance with what we expect due to heating effects.  The best sinusoidal fit to the data has an
amplitude of $0.27\pm0.18$ (solid line in Figure~\ref{f8}).
This sinusoidal fit to the data implies that the secondary star contributes
$\sim39\%$ of the K-band flux at phase 0.0, while at phase 0.5 this percentage is reduced to
$\sim15\%$.  Hence, the absorption in the hemisphere of the secondary star
nearest to the primary star is $\sim0.4$ times the strength of the absorption in
the hemisphere facing away from the primary.

Averaged over orbital phase these results suggest that the secondary star
contributes $29\pm13\%$ of the K-band flux.  This is considerably different from the $65\pm5\%$
estimated by Dhillon \ea (2000).  The AAVSO quick look light curves
of RW Tri at the time of our UKIRT observations show that RW Tri was at a magnitude of V$\sim13$, but during the observations of Dhillon \ea
(2000) RW Tri appeared to be in a low state with a magnitude of
$\sim13.8$. Thus, RW Tri was approximately a factor of 2 brighter during our
observations than when observed by Dhillon \ea (2000), accounting for the
different estimates of the secondary star contribution. Assuming that the flux
of the secondary star is constant, the accretion disc and
stream in RW Tri increased by a factor of $\sim 4$ in brightness, between 
the observations of Dhillon \ea (2000) and ours.

\subsection{Secondary Star Rotational Velocity}
\begin{figure}
\centering
\vspace*{6.0cm}
\leavevmode
\includegraphics{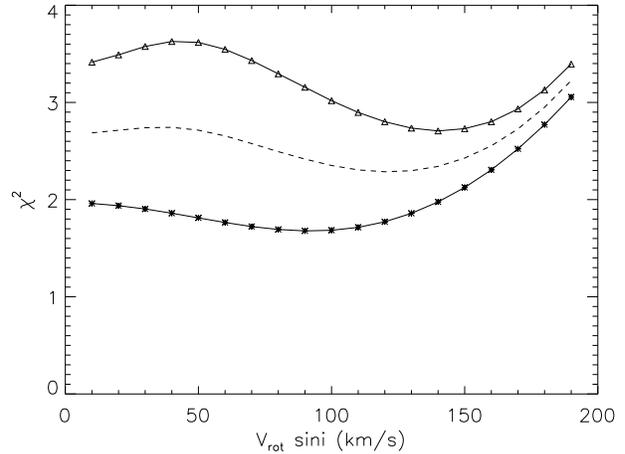}
\caption{Plot of $\chi_{\nu}^{2}$ obtained using the average of all templates
after artificially broadening by $V_{rot}\sin i$ in the range
$10-200$km/s. This shows that the
minimum $\chi_{\nu}^{2}$ for night 1 ($\ast$), night 2 ($\bigtriangleup$) and
the average of the two nights night (dashed line)
are 90km/s, 140km/s and 120km/s respectively.}
\label{f7a}
\end{figure}

We noted previously that the secondary star features in RW~Tri are significantly
broader than in the template spectra. This is likely to be
due to broadening caused by the rotation of the (phase-locked) secondary star
as it orbits the white dwarf.  
The rotational velocity of the secondary star can be estimated by
artificially broadening the template star spectrum which is assumed to have low
$V_{rot}\sin i$, and fitting it to
the RW Tri spectra. The continuum was removed from the RW Tri and
template star spectra using a low order polynomial, after masking out strong absorption
features. Each template was then artificially broadened in the velocity range $V_{rot}\sin i
=10-200$km/s in steps of 10km/s, assuming partial limb darkening (linear limb
darkening coefficient = 0.5, North \ea 2000).  These broadened 
template spectra were then compared with the orbital velocity corrected 
RW Tri spectra on each night. Residual spectra
were produced by subtracting a constant times the shifted broadened template from each RW Tri spectrum; the constant was adjusted to minimise the scatter on
each residual spectrum.  A boxcar average smoothing was applied to the
residual spectrum to eliminate any large scale structure. The reduced
chi-squared ($\chi_{\nu}^2$) was calculated between each residual and smoothed
spectrum in the wavelength regions containing the Na I absorption feature (22040\AA\ to
22108\AA) and the Ca I absorption feature (22614\AA\ to 22689\AA).  The average results can be seen
in Figure~\ref{f7a} where we plot night 1 and night 2 separately, and also
combined.

The best fit (minimum $\chi_{\nu}^{2}$) for night 1 and 2 is obtained with $V_{rot}\sin i
\sim 90$km/s and $V_{rot}\sin i \sim 140$km/s respectively (Figure~\ref{f7a}).  Based on
 the $\chi_{\nu}^2$ distribution in Figure~\ref{f7a}, we estimate a mean
$V_{rot}\sin i=120\pm20$km/s.  All template stars gave the same order of
Chi-squared values, confirming again that the data are not template sensitive in this wavelength band.  The minimum
$\chi_{\nu}^2$ values had a range over all the templates of $V_{rot}\sin i$
from $80-100$km/s for night 1, and $130-150$km/s for night 2.  We confirmed that
the intrinsic $V_{rot}\sin i$ of each template was consistent with 0km/s, by cross correlating the
templates against each other.

The analysis of the orbital radial velocity was not changed significantly when
we broadened the template lines by 120km/s. This is because rotational broadening affects the profile of the absorption lines, but not the position of the line centroids which govern the cross correlation peak positions.

\section{Discussion}
To calculate the masses of the two component stars in the system we can use the mass
ratio ($q$),
\begin{equation}
q=\frac{K_1}{K_2}=\frac{M_2}{M_1},
\label{e6}
\end{equation}
where {\it $K_1$} and {\it $K_2$} are the radial velocity amplitudes of the
primary and secondary stars respectively, and {\it $M_1$} and {\it $M_2$} are
the primary and secondary masses respectively.

\subsection{Primary Star Velocity Measurements}
There have been a number of different estimates of $K_1$ based on measurements of the
optical emission lines in RW~Tri and these have yielded
a
range of values.  Doppler maps of RW Tri (Kaitchuck \ea 1983) indicate that the 
He II $\lambda$4686\AA\
emission arises from the inner accretion disc region, He I $\lambda$4471\AA\ emission is found 
further out in the accretion disc, and H$\beta$ and H$\gamma$ emission originates
from the outer regions of the accretion disc and the inner face of the secondary
star.

Still \ea (1995) measured the emission line centroids and obtained $K_1$
values of $208\pm8$ km/s from H$\beta$, $223\pm11$ km/s from H$\alpha$, 
and $216\pm9$ km/s from He II ($\lambda$4686\AA).  
The He II measurement of Still \ea (1995) is consistent with the $197\pm20$ km/s measured by
Kaitchuck \ea (1983), who also found a $K_1$ velocity of $\sim170\pm20$ km/s for He I.  
Still \ea (1995) also used
the convolution technique of Schneider \& Young (1980) to measure the wings of the
H$\beta$, H$\gamma$ and He II emission lines. The line wings, which come from high 
velocity gas in the inner disk, are in principle more likely to reflect the motion of 
the white dwarf than the line cores. The latter may well be contaminated
by emission from the secondary star, the accretion stream, and the bright spot. 
However, Still \ea found that the measured velocities for the
emission line wings in RW~Tri were inconsistent with the  velocities obtained from 
Doppler maps.  These
inconsistencies could be due to absorption affecting the wings of the
accretion disc emission lines.

Recently HST data have been used to measure the velocity of narrow absorption
features in the UV.
Mason \ea (in prep.) detected velocity shifts in the UV absorption lines and found they had
the same orbital phase as that expected for the white dwarf.  The lines appear
to originate in
a layer above the inner accretion disk, and their motion may therefore mirror that of 
the white dwarf.  By cross correlating the average spectrum through the orbital cycle 
they find $K_1=296\pm5$ km/s.

\subsection{Effects on the Secondary Star Radial Velocity Amplitude}
The apparent value of $K_2$ may be modified by  heating, line-quenching,
and line contamination (Friend \ea 1990). {\it Heating} of the secondary
star occurs due to hard photons from the accretion disc. {\it Line quenching}
occurs due to the ionisation of the absorption lines by flux from the disc. Both
irradiation and line quenching can deplete the absorption line strength on the
surface of the secondary star that faces the accretion disc, and 
shift the apparent centroid of the absorption line region to the hemisphere facing away from the
disc. This would lead to an over estimate of the $K_2$ value. The magnitude of this effect 
may be reduced however if the accretion disk has a thick rim which shields the secondary. 
This appears to be the case in RW~Tri (Mason, Drew \& Knigge 1997). {\it Line
contamination} may occur due to weak disc features. This could lead to
variations in the line strengths of the absorption through the orbit. 

To account for these effects, Wade \& Horne
(1988) estimated the likely correction required for the radial velocity amplitude
of the secondary star.  This ``K-correction'' is given by,
\begin{equation}
\Delta K=\frac{\Delta R}{a_2}K_2=\frac{fR_2}{a_2}K_2,
\label{extra2}
\end{equation}
where $\Delta${\it R} is the displacement between the effective centre and the
centre of mass of the secondary star, $R_2$ is the secondary star radius,
$|{\it f}|<$1 is a weighting factor representing the strength of the absorption
feature, and $a_2$ is the distance of the centre of mass of the secondary star
from the centre of mass of the system given by,
\begin{equation}
a_2=\frac{a}{1+q},
\end{equation}
where $a$ is the separation of the component stars and $q$ is the mass ratio
($q=M_2/M_1$). 

To estimate the K correction, we use
$f =4/5\pi\sim0.25$ where the front hemisphere of the secondary star has $\sim0.4$ of the
absorption of the back hemisphere.  This is based on the measured change in the
amplitude of the absorption features with orbital phase (Section
3.4).  Each hemisphere is assumed to have uniform absorption, and we relate
$R_2/a$ to the mass ratio $q$ using
\begin{equation}
\frac{R_2}{a}=\frac{0.49q^{2/3}}{0.6q^{2/3}+\ln(1+q^{1/3})}
\label{eadded}
\end{equation}
(Eggleton, 1983).  

\subsection{Mass Ratio Theory}
Consider the conservative mass transfer equation,
\begin{equation}
\frac{\dot{R_L}}{R_L}=\frac{2\dot{J}}{J}+\frac{2(-\dot{M_2})}{M_2}\left(\frac{5}{6}-q\right),
\label{e7}
\end{equation}
where R$_L$ is the radius of the Roche Lobe, $J$ is the orbital angular
momentum, and -$\dot{M_2}$ is the
instantaneous mass transfer rate (Frank, King \& Raine 1992). 

When $q(=M_2/M_1)<5/6$ then $\dot{R_L}>0$, so the Roche lobe
expands, reducing the mass transfer, and the system is stable. In order to sustain long lived mass
transfer the secondary star must expand in size relative to the Roche lobe, otherwise the lobes
detach from the star and mass transfer stops. Evolution of the secondary star
is one possibility, but for the secondary to evolve within the age of the
Galaxy, it must be spectral type G0 or earlier (Patterson 1984). Most CV secondaries have spectral types later than G0, so a more
likely solution for stable mass transfer is angular momentum loss due to either gravitational radiation
and/or magnetic braking.  The loss of angular momentum shrinks the binary
system therefore enabling sustained mass transfer to occur. 

When $q>5/6$ then $\dot{R_L}<0$, and
the Roche lobe shrinks. Mass transfer will
therefore increase, and the system
will become unstable unless the secondary star can contract rapidly enough to keep its
radius smaller than the radius of the Roche lobe. If the
secondary star obeys the main sequence mass-radius relation $R_2\propto M_2$, and the
radius of the star responds to changes in its mass on a thermal time scale, Equation~\ref{e7} becomes,
\begin{equation}
-\frac{\dot{J}}{J}=-\frac{\dot{M}_2}{M_2}\left(\frac{4}{3}-q\right),
\label{e8}
\end{equation}
yielding a critical upper mass ratio ($q_{crit}$). When $q>4/3$ the secondary
star will not shrink rapidly enough to keep pace with the Roche lobe. 
There will be a spontaneous overflow and mass transfer becomes unstable.

The secondary star in a CV is a late type low mass star with a deep 
convective envelope, and therefore
loses mass on a dynamical time scale governed by the star's adiabatic response.  Considering a complete
polytrope with a polytropic index of $n=3/2$ (Hjellming \& Webbink 1987), the
mass-radius relation for the secondary star becomes $R_2\propto
M_2^{-1/3}$, and hence Equation~\ref{e7} becomes,
\begin{equation}
-\frac{\dot{J}}{J}=-\frac{\dot{M}_2}{M_2}\left(\frac{2}{3}-q\right),
\label{e9}
\end{equation}
producing a lower mass ratio limit ($q_{ad,fc}$). When $q > 2/3$ the star can not remain within its Roche lobe in hydrostatic
equilibrium, and mass transfer occurs on dynamical time scales.  When $q < 2/3$,
the star becomes stable on a dynamical time scale and mass transfer occurs due
to the slow expansion of the star via nuclear evolution or angular momentum loss
causing the Roche lobe to contract. 

The secondary star in RW Tri may
not be fully convective so the true adiabatic mass ratio will be higher.  In
the case where the secondary star has a convective envelope, but a radiative
core, the mass-radius relation becomes $R_2\propto M_2^{1/3}$, leading to a mass
ratio of $q_{ad,rc}=1$ for the adiabatic response (Hjellming \& Webbink 1987). 

\subsection{Mass Ratio}
We first calculate the mass
ratio of RW Tri using the various estimates of the component star radial velocity
amplitudes. The most reliable estimate for the secondary star velocity is from
the K-band data because there is
not enough detail in the I-band data to be sure that they are not affected by
telluric lines and background emission etc. When combined with the K-band secondary star velocity
($221\pm29$km/s), the various estimates of the primary star velocity amplitude 
discussed in Section 4.1
lead to a range of mass ratios of $0.8-1.3$ as expressed in Table~\ref{t3} (column 4). 

\begin{table}
\caption{Mass ratio results using our best estimate of the secondary star velocity, with a combination of white dwarf velocities.}
\begin{tabular}{ccccc}
\hline
Feature & $K_1$  & $K_2$  & $q$ using & $q$ using \\ 
&(km/s) & (km/s)& $K_2$  & $\Delta K\sim24\%$ \\
\hline
UV  & $296\pm5$ & $221\pm29$ & $1.34\pm0.18$ & $1.66\pm0.27$\\
H$\alpha$ & $223\pm11$ & $221\pm29$ & $1.01\pm0.14$ & $1.25\pm0.21$\\
He II & $216\pm9$ & $221\pm29$ & $0.98\pm0.14$ & $1.21\pm0.20$\\
H$\beta$ & $208\pm8$ & $221\pm29$ & $0.94\pm0.13$ & $1.17\pm0.20$\\
He I & $170\pm20$ & $221\pm29$ & $0.77\pm0.14$ & $0.96\pm0.19$\\
\hline
\end{tabular}
\label{t3}
\end{table}

The $K_1$ velocity values that {\it a priori} are most likely to reflect the motion of the white dwarf are
the UV absorption lines of Mason \ea (2002), and the He II emission lines of
Still \ea (1995), because they both originate in regions close to the white
dwarf.  These velocities therefore give us a `most likely' mass ratio in the
range $1.0-1.3$. To consider the effects of the ``K-correction'' on our most likely mass ratio
range, we use Equation~\ref{extra2} and~\ref{eadded}, and $f\sim0.25$ with
$q=1.0-1.3$, which corresponds to a range in $\Delta K$ of $\sim 19\%$ to $\sim24\%$.
Thus after applying the most likely value for the
secondary star heating the value of $K_2$ in RW Tri is $\sim178$km/s, implying
a revised mass ratio, $q$, in the range $1.2-1.7$ (Table~\ref{t3}, column 5). 

\begin{figure}
\centering
\vspace*{6.0cm}
\leavevmode
\includegraphics{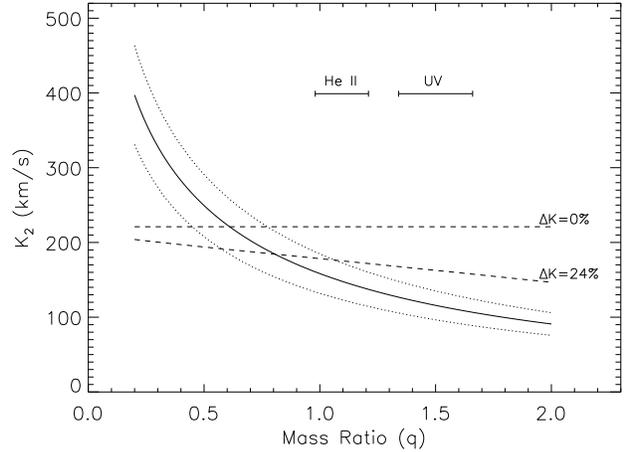}
\caption{Radial velocity of the secondary star ($K_2$) versus a range of mass
ratios ($q=M_2/M_1$).  The solid curved line represents $K_2$ calculated using
$V_{rot}\sin i=120$km/s, and Equations~\ref{eaddedd} and~\ref{eadded}. The
dotted curved lines represent $K_2$ using $V_{rot}\sin i=120\pm20$km/s.  The
dashed straight lines represent the variation of K-corrected $K_2=221$km/s
(Section 4.2)as a function of mass ratios, when $\Delta K=0\%$ and $\Delta
K=24\%$. The horizontal bars labeled He II and UV represent the mass
ratio range of He II and UV respectively, using $\Delta K=0\%$ and
$\Delta K=24\%$ from Table~\ref{t3}.}
\label{f7b}
\end{figure}

Alternatively we can calculate the mass ratio of RW Tri using the rotational
broadening of the secondary star, independent of $K_1$.  Assuming that the
secondary star rotates in phase with the binary orbit we use,
\begin{equation}
\frac{K_2}{V_{rot}\sin i}=\left[(1+q)\left(\frac{R_2}{a}\right)\right]^{-1}
\label{eaddedd}
\end{equation}
where $R_2/a$ is found using
Equation~\ref{eadded}. The results are shown in Figure~\ref{f7b} where the
solid line represents the $K_2$ values implied by $V_{rot}\sin i=120$km/s (from Section 3.6) as a
function of mass
ratio between $q=0.2$ and $q=2.0$.  The dotted lines show the effects of
changing $V_{rot}\sin i$ by $\pm20$km/s. The dashed lines in Figure~\ref{f7b}
represent the K-corrected secondary star velocity as a function of the mass
ratio, for the cases $\Delta K=0\%$ and $\Delta K=24\%$.  We find a self-consistent value of $q$ in
the range $0.5-0.8$ when no correction for possible heating effects is applied
($\Delta K=0\%$). This does not agree with the mass
ratio range derived for either He II or UV $K_1$ velocities, which are indicated in 
Figure~\ref{f7b}. When $\Delta
K=24\%$, the allowed mass ratio range is $q=0.6-1.1$ (Figure~\ref{f7b}), and
again does not overlap with either the He II or  UV range. 

\begin{figure}
\centering
\vspace*{6.0cm}
\leavevmode
\includegraphics{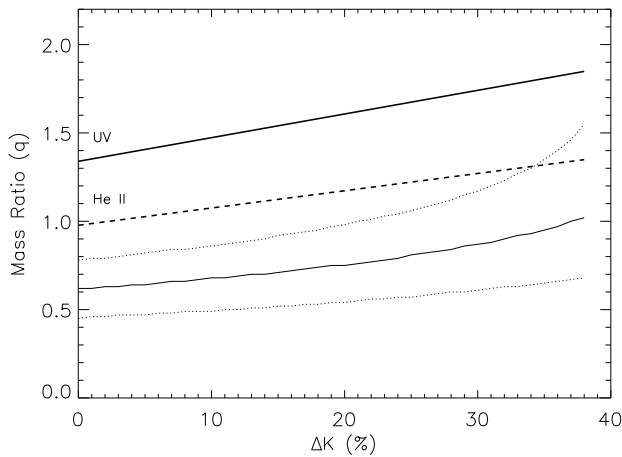}
\caption{The possible mass ratio ranges of RW Tri ($q=M_2/M_1$) as
a function of the ``K-correction'' of the secondary star ($\Delta K$) due to
heating effects.  The solid black curved line represents the mass ratio ($q$)
found when a $K_2$ using $V_{rot}\sin i=120$km/s (solid black curved line in
Figure~\ref{f7b}) coincides with a K-corrected $K_2=221$km/s (dashed lines
in Figure~\ref{f7b}), as a function of $\Delta K$. The
dotted curved lines represent the mass ratio as a function of $\Delta K$ using
$V_{rot}\sin i=120\pm20$km/s. The thick solid and thick dashed lines represent the mass
ratio range of He II and UV respectively, as a function of $\Delta K$ using Equation~\ref{extra2}.}
\label{f7c}
\end{figure}

This can be more clearly seen in
Figure~\ref{f7c} where the derived values of $q$ are expressed as a
function of $\Delta K$.  The solid black line in
Figure~\ref{f7c} shows the mass
ratio at which the $K_2$ value implied by $V_{rot}\sin i=120$ km/s equals the
K-corrected value of $K_2$ (adopting an observed value of $K_2$ of
221 km/s). Again the dotted lines
indicate the effect of changing $V_{rot}\sin i$ by $\pm 20$ km/s. The dashed
and thick solid lines represent the mass ratio derived using the He II and UV line
estimates of $K_1$ in combination with the K-corrected value of $K_2$, 
as a function of
$\Delta K$. Figure~\ref{f7c} shows that $V_{rot}\sin i$ is consistent
with the He~II-based mass ratio range only for $\Delta K>34\%$, and
higher still for the UV-based $K_1$ value (cf. our best
estimate of $\Delta K\sim24\%$).
This is greater than the
we find in Section 4.2, suggesting that velocities found
using the UV and He II emission lines may contain non-orbital components.

\subsection{Masses}
Using the Roche Lobe geometry, a relationship between the mass ratio ($q$), 
orbital inclination angle ($i$) and eclipse duration was derived by Chanan
\ea (1976) and also by Horne (1993).  Because the temperature of the accretion disk
increases towards its centre, the UV emission of the disk will be more concentrated around the
white dwarf than the optical emission. Thus an eclipse width at half light measured in the UV
is likely to be a better approximation to the white dwarf 
eclipse duration than one measured in the optical band. Using the
relationships of Chanan \ea (1976) and Horne (1993), the mass ratio, and the eclipse
width at half light of $0.077\pm0.002$ from the UV light curves of
Mason \ea (1997), a set of inclination angles can be calculated.  These inclination angles
range from $73^o$ to $79^o$ for mass ratios between $0.5$ and $1.1$. 
Hence a range of masses for
the primary and secondary stars can be derived using,
\begin{equation}
M_1\sin^3i=\frac{P_{orb}K_2}{2\pi G}[K_1+K_2]^2,
\label{ee1}
\end{equation}
and,
\begin{equation}
M_2\sin^3i=\frac{P_{orb}K_1}{2\pi G}[K_1+K_2]^2.
\label{ee2}
\end{equation}

\begin{figure*}
\centering
\vspace*{10.0cm}
\leavevmode
\includegraphics{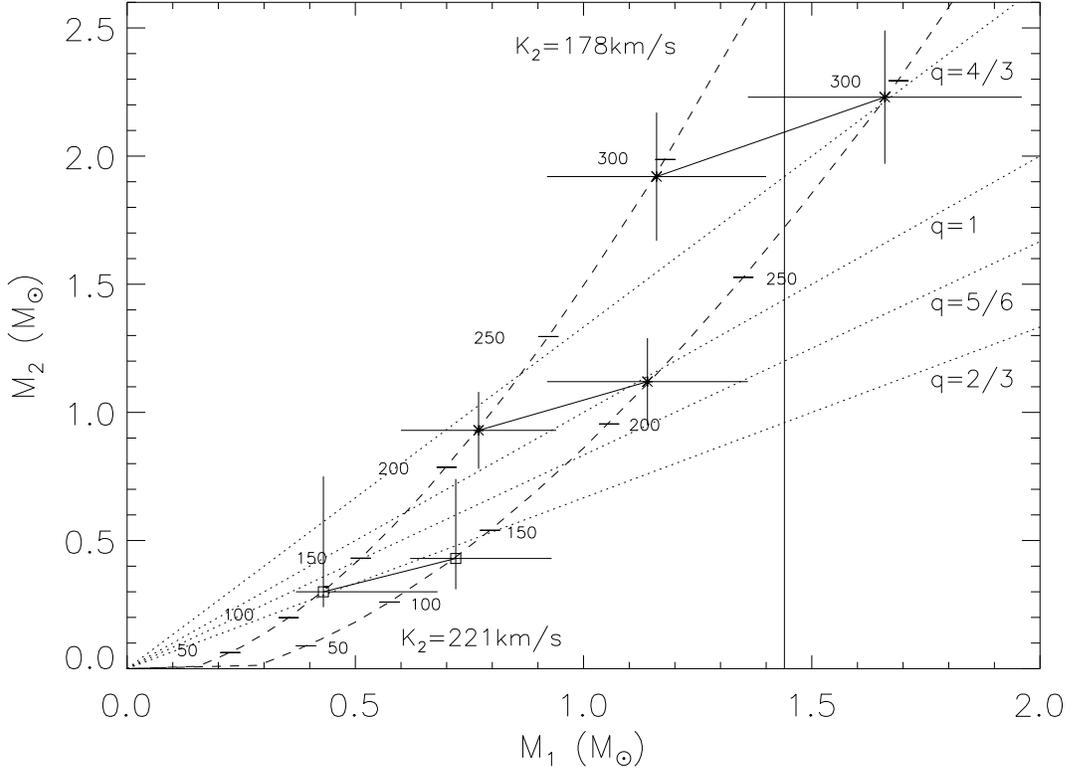}
\caption{The mass of the component stars in RW~Tri, derived by combining the
measurement of $K_2$ with different estimates of $K_1$ (represented as $\ast$) , and from the combination
of $K_2$ with $V_{rot}sin i$ (represented by $\Box$). In each case we show values 
corrected for the effects of heating of the secondary using $\Delta K\sim 24\%$
($K_2=178$ km/s - Section 4.3) and also with no heating correction ($K_2=221$ km/s),
linked by a straight line. The upper set of results are obtained using the UV
absorption $K_1=296\pm5$ km/s velocity (Mason \ea in prep), and the middle set 
of results are obtained using the He II accretion disc emission 
$K_1=216\pm9$ velocity (Still \ea 1995).  
The error bars combine the
uncertainties in $K_1$ and $K_2$. The lower set of results show the mass values
derived from the $K_2$ measurement combined with
$V_{rot}\sin i=120\pm20$km/s.  The error
bars are calculated using the $\pm20$km/s uncertainty on $V_{rot}\sin i$. 
The vertical solid line represents the $1.44M_{\odot}$
Chandrasekhar mass limit for white dwarf stars. The two dashed lines represent
the locus of 
masses calculated when $K_2=221$km/s and $K_2=178$km/s respectively with
varying $K_1$ values; the tick marks on the two dashed lines mark $K_1$
velocity values of 50km/s, 100km/s, 150km/s, 200km/s, 250km/s and 300km/s.  The four dotted lines represent the
$q_{crit}=4/3$ thermal mass transfer limit, $q_{ad,rc}=1$ adiabatic response
with a radiative core, $q=5/6$ and $q_{ad,fc}=2/3$ fully convective lower adiabatic
response limit respectively (Section 4.2).}
\label{f10}
\end{figure*}

The results derived using the UV and He II $K_1$ velocities are shown in
Figure~\ref{f10}. Using the UV measurement, both the uncorrected and
K-corrected mass ratios lie above the critical value
of 4/3 (Section 4.3), in the region of the diagram where mass transfer is
unstable.  The upper mass for the primary star exceeds the Chandrasekhar mass
limit of $1.44M_{\odot}$, and the secondary star mass is also very large and
inconsistent with that of a main sequence star.  This reinforces our suspicions that 
the UV velocities contain a
non-orbital component. A reduction of $K_2$ below the K-corrected
value of 178km/s would decrease the
secondary mass, but further increase the mass ratio.

Adopting instead the $K_1$ measurement derived from He II data, we find values of $q$
that broadly lie between $q_{crit}=4/3$ and $q_{ad,rc}=1$. The masses for the primary and
secondary stars lie in the range of $0.8-1.1M_{\odot}$, and $0.9-1.1M_{\odot}$ respectively. 
The white
dwarf mass is well within the Chandrasekhar mass limit of $1.44M_{\odot}$, and
includes the mean primary mass of $0.8M_{\odot}$ for CV systems with
$P_{orb}>3$ hrs (Smith \& Dhillon 1998b).  The lower limit for the
secondary star mass, however, exceeds the predicted mass of $0.55 M_{\odot}$
based on the main sequence mass-radius relation for an orbital period of 5.25 hours
(Echevarr\'{\i}a 1983). 

Given the evidence for a non-orbital component in the UV absorption line
velocities, we cannot be certain that the optical emission line velocities are not
similarly affected.  If we force the secondary star mass to its equivalent main
sequence value, $0.55M_{\odot}$, we would predict a primary star velocity of
$152-169$km/s  for $K_2=221-178$km/s.  This is closer to the $K_1$ value of
$170\pm20$km/s derived by Kaitchuck \ea
(1983) from the He I emission line. The implied mass ratio for a $K_1$ velocity
of 170km/s is in the range 0.8 ($\Delta K=0\%$) to 1.0 ($\Delta K=24\%$),
and is consistent with the rotational velocity of the secondary that we derive (Figure~\ref{f7c}). 

Alternatively, we can obtain mass values using the combination of $V_{rot}\sin i$ and $K_2$
without any assumptions about $K_1$. 
Using the mass ratio values of $0.6-1.1$ for
$\Delta K=24\%$, and $0.5-0.8$ for
$\Delta K=0\%$ from Section 4.4, and assuming $K_2$ values of
178km/s and 221km/s respectively, the stellar masses can be calculated using Equations
~\ref{e6},~\ref{ee1}, and~\ref{ee2}. Figure~\ref{f10} shows the results.
The most likely mass of the primary and
secondary stars lies between $0.4-0.7M_{\odot}$ and $0.3-0.4M_{\odot}$
respectively for the range of $\Delta K=24-0\%$, and better agree with the expected 
masses for the
component stars in a CV. The expected value of $K_1$ based on these results is
in the range 120-130 km/s, with an upper limit of approximately 190 km/s. 

\section{Conclusions}
I-band observations of RW Tri yield a secondary star radial velocity amplitude of
$250\pm47$ km/s using the skew mapping technique. K-band observations of RW Tri
provide us with a secondary star velocity of $221\pm29$ km/s which is
obtained directly from the observations without using complex
mapping methods. The two velocities are consistent within the errors. 

We estimate the rotational velocity of the secondary star to be $120\pm20$km/s 
using the K-band UKIRT observations. Combining this velocity
with the secondary star radial velocity corrected for non-uniform heating
derived from the variation in the
absorption line strengths, we find a mass ratio range of $0.6-1.1$, which
contains the lower adiabatic response limit ($q_{ad,fc}=2/3$). The
uncorrected radial velocity results lead to a mass ratio range of $0.5-0.8$. 

Combining the radial velocity amplitudes of the secondary star with the radial 
velocity of the primary determined from  He II
emission lines in the optical (Still \ea 1995) and narrow absorption lines in the UV 
(Mason \ea in prep.), yields
a range of mass ratios of $1.0-1.3$ and $1.2-1.7$ respectively, based on a
range in the secondary heating correction of $\Delta K=0-24\%$. The UV mass
ratio range lies above the critical mass ratio ($q_{crit}=4/3$); this range is also
inconsistent with the rotational velocity results, indicating that the UV
velocity very likely includes a non-orbital component.  The He II mass ratio range 
lies between $q_{crit}=4/3$ and q=5/6, but is only marginally consistent with 
the measured secondary rotational velocity, and may also contain a non-orbital
component. By combining the data on the 
rotational broadening of the secondary with its measured orbital velocity, with no 
assumptions regarding the white dwarf velocity,
we find most likely values of the primary and secondary masses that lie in
the range
$0.4-0.7M_{\odot}$ and $0.3-0.4M_{\odot}$ respectively, depending on the degree of
secondary star heating. The most likely value of $K_1$ is predicted to be 120-130 km/s.

\section*{Acknowledgments}
We thank Paul Hirst for help with CGS4 data reduction. The United
Kingdom Infra-Red Telescope is operated by the Joint Astronomy Centre on behalf
of the U.K. Particle Physics and Astronomy Research Council. Also thanks to Dave
Vande Putte for valuable discussion on the INT data, and Tom Marsh for use of,
and help with, MOLLY. In our research, we have used, and acknowledge with
thanks, data from the AAVSO International Data base, based on observations
submitted to the AAVSO by variable star observers worldwide. We would also like
to thank the referee for a careful reading of the manuscript and useful comments.

\end{document}